\documentclass[11pt, a4paper]{article}
\usepackage[margin = 1in]{geometry}
\usepackage{libertinus}
\usepackage{amsmath}
\usepackage{amssymb}
\usepackage{amsfonts}
\usepackage{amsthm} 
\usepackage{graphicx}
\usepackage{comment}
\usepackage{thm-restate}
\usepackage{hyperref}
\usepackage{cleveref}
\usepackage{here}
\usepackage{cite}
\usepackage{enumerate}
\usepackage{tabularx}
\usepackage{environ}
\usepackage{url}
\usepackage[normalem]{ulem}
\usepackage{color}
\usepackage[T1]{fontenc}
\usepackage{booktabs}

\newtheorem{theorem}{Theorem}
\newtheorem{lemma}[theorem]{Lemma}
\newtheorem{fact}[theorem]{Fact}
\newtheorem{corollary}[theorem]{Corollary}
\newtheorem{claim}[theorem]{Claim}
\newtheorem{observation}[theorem]{Observation}

\usepackage{algorithm}
\usepackage[noend]{algpseudocode}

\algrenewcommand\algorithmicrequire{\textbf{Input:}}
\algnewcommand\algorithmiccontinue{\textbf{continue}}
\algnewcommand\algorithmicbreak{\textbf{break}}

\global\long\def\E{\mathbb{E}}%
\global\long\def\M{\mathcal{M}}%
\global\long\def\I{\mathcal{I}}%
\global\long\def\tO{\widetilde{O}}%
\global\long\def\TO{\tilde{O}_{\varepsilon}}%
\global\long\def\eps{\varepsilon}%
\global\long\def\epsilon{\varepsilon}%
\global\long\def\tO{\tilde{O}}%
\global\long\def\eps{\varepsilon}%

\newcommand{\Poly}{\mathrm{poly}}

\newcommand{\ds}{\displaystyle}

\newcommand{\N}{\mathcal{N}}
\newcommand{\NL}{\mathcal{N}'}
\newcommand{\ML}{\mathcal{M}'}
\newcommand{\IL}{\mathcal{I}'}
\newcommand{\OPT}{\mathrm{OPT}}
\newcommand{\fmax}{f_{\max}}
\newcommand{\Tmax}{\Theta_{\max}}
\newcommand{\Tmin}{\Theta_{\min}}
\newcommand{\tmax}{t_{\max}}

\newcommand{\hg}{\widehat{\gamma}}
\newcommand{\hla}{\widehat{\lambda}}
\newcommand{\Binit}{B_{\mathrm{init}}}
\newcommand{\Est}{\textnormal{\textsc{Estimate}}}
\newcommand{\Phase}{\textnormal{\textsc{Phase}}}

\let\widetilde\tilde

\newcommand{\ShowComment}{true}
\ifdefined\ShowComment
\newcommand{\tatsuya}[1]{{\bf \color{red} tatsuya: #1}}
\else
\newcommand{\tatsuya}[1]{}
\fi

\title{Nearly-Linear Query Algorithm for Submodular Maximization subject to a Matroid Constraint}
\author{Tatsuya Terao\thanks{
Research Institute for Mathematical Sciences, Kyoto University.
E-mail: ttatsuya@kurims.kyoto-u.ac.jp
}
}

\date{}

\begin{document}

\maketitle

\begin{abstract}
We consider monotone submodular maximization subject to a matroid constraint. The seminal continuous greedy algorithm of Calinescu, Chekuri, P\'{a}l, and Vondr\'{a}k~[SICOMP'11] achieves the tight approximation of $(1-1/e)$. Since then, a line of works by Badanidiyuru and Vondr\'{a}k~[SODA'14], Buchbinder, Feldman, and Schwartz~[SODA'15 \& MOR'17], Kobayashi and Terao~[ICALP'24], Buchbinder and Feldman~[FOCS'24 \& SICOMP'25], and Buchbinder and Feldman~[STOC'25] has developed efficient $(1-1/e-\varepsilon)$-approximation algorithms for this problem.

We present a combinatorial randomized $(1-1/e-\varepsilon)$-approximation algorithm that requires $\tilde{O}_{\varepsilon}(n)$ independence-oracle and value-oracle queries, where $n$ is the number of elements in the matroid. This is the first $(1-1/e-\varepsilon)$-approximation algorithm with both independence-oracle and value-oracle query complexities nearly-linear in $n$ and polynomial in $1/\varepsilon$. To achieve this, we develop a new local search algorithm, which we call \emph{protected local search}, based on the potential function of Buchbinder and Feldman~[FOCS'24 \& SICOMP'25].
\end{abstract}

\newpage

\section{Introduction}

Submodular maximization is a fundamental and well-studied problem in theoretical computer science and combinatorial optimization. 
In particular, the problem of maximizing a nonnegative monotone submodular function subject to a matroid constraint is one of the most extensively investigated settings.

The study of submodular maximization was initiated by a seminal work of Fisher, Nemhauser, and Wolsey in the 1970's \cite{nemhauser1978analysis, fisher1978analysis, nemhauser1978best}.
They showed that, for monotone submodular maximization, the natural greedy algorithm achieves a $(1 - 1/e)$-approximation for a cardinality constraint and a $\frac{1}{2}$-approximation for a matroid constraint.
They~\cite{nemhauser1978best} also showed that, for any constant $\delta > 0$, achieving a $(1-1/e+\delta)$-approximation requires exponentially many value-oracle queries, even under a cardinality constraint.
Thus, the factor $1-1/e$ is optimal for a cardinality constraint.

To obtain an optimal $(1 - 1/e)$-approximation algorithm for a matroid constraint, Calinescu, Chekuri, P\'{a}l, and Vondr\'{a}k~\cite{calinescu2011maximizing} developed a new framework based on continuous optimization and a rounding technique.
Their algorithm first obtains a $(1-1/e)$-approximate solution to the problem of maximizing the multilinear extension of $f$, a natural continuous extension of $f$,  via the continuous greedy algorithm, and then rounds it to an integral solution via pipage rounding~\cite{ageev2004pipage}.
Consequently, their algorithm achieves the optimal $(1 - 1/e)$-approximation.
Note that, although their algorithm runs in polynomial time, its running time is very high. 

Since submodular maximization has a number of applications, providing efficient approximation algorithms is a fundamental task both in theory and practice. 
Thus, it has received considerable attention to develop fast submodular maximization algorithms  that achieve an approximation close to the optimal factor, typically with an approximation factor of $1 - 1/e - \varepsilon$.

In submodular maximization subject to a matroid constraint, it is standard to assume that the objective function $f$ is accessed through a value oracle and that the family ${\cal I} \subseteq 2^\N$ of independent sets is accessed through an independence oracle.
In this setting, the efficiency of an algorithm is usually measured by the number of queries it makes to these oracles.

Badanidiyuru and Vondr\'{a}k~\cite{badanidiyuru2014fast} presented a fast algorithm that achieves an almost optimal approximation factor $1 - 1/e - \varepsilon$, for any $\varepsilon > 0$, for a matroid constraint.
Their algorithm uses $\tilde{O}_{\varepsilon} \left( nr \right)$ value-oracle queries and $\TO\left( n + r^2 \right)$ independence-oracle queries\footnote{The $\tilde{O}$ notation hides polylogarithmic factors in $n$, and the $\tilde{O}_{\varepsilon}$ notation additionally hides polynomial factors in $\varepsilon^{-1}$.}, where $n$ is the number of elements in the matroid and $r$ is the rank of the matroid.
Their algorithm combines a fast implementation of the continuous greedy algorithm, which uses $\TO(nr)$ value and $\TO(n)$ independence oracle queries, with the swap rounding of Chekuri, Vondr\'{a}k, and Zenklusen~\cite{chekuri2010dependent}, which rounds the resulting convex combination of $1/\varepsilon$ bases using $O(r^2/\varepsilon)$ independence oracle queries.

Buchbinder, Feldman, and Schwartz~\cite{buchbinder2017comparing} developed another fast implementation of the continuous greedy algorithm, which exhibits a trade-off between the numbers of value and independence oracle queries.
In particular, they obtained a $(1 - 1/e - \varepsilon)$-approximation algorithm using $\TO(n\sqrt{r})$ value oracle queries and $\TO(n\sqrt{r} + r^2)$ independence oracle queries.
As in the algorithm of Badanidiyuru--Vondr\'{a}k, the $\TO(r^2)$ term in the independence query complexity is due to swap rounding.

Kobayashi and Terao~\cite{kobayashi_et_al:LIPIcs.ICALP.2024.100} improved the independence query complexity of the swap rounding of Chekuri, Vondr\'{a}k, and Zenklusen~\cite{chekuri2010dependent} to $\TO(r^{3/2})$.
Combined with the algorithm of Buchbinder, Feldman, and Schwartz, this yields a $(1 - 1/e - \varepsilon)$-approximation algorithm using $\TO(n\sqrt{r})$ value and independence oracle queries.

In another line of work, Filmus and Ward~\cite{filmus2014monotone} presented the first combinatorial $(1-1/e)$-approximation algorithm, which requires no rounding.
Their algorithm is based on \emph{non-oblivious local search}, i.e., local search guided by an auxiliary potential function instead of $f$, which is estimated by sampling and thus makes the algorithm randomized.

All of the above $(1-1/e-\varepsilon)$-approximation algorithms for a matroid constraint are randomized.
In a breakthrough work, Buchbinder and Feldman~\cite{buchbinder2025deterministic} presented the first deterministic $(1-1/e-\varepsilon)$-approximation algorithm, which is based on the non-oblivious local search and uses $\widetilde{O}(nr \cdot 2^{O(1/\varepsilon)})$ value oracle queries and $\TO(nr)$ independence oracle queries.
With randomization, these query complexities can be reduced to $\widetilde{O}((n + r\sqrt{n}) \cdot 2^{O(1/\varepsilon)})$ and $\TO(n + r\sqrt{n})$, respectively.

Subsequently, Buchbinder and Feldman~\cite{buchbinder2025extending} introduced the extended multilinear extension, which enables deterministic implementations of both the fractional maximization step and the rounding step of the continuous greedy framework.
As a by-product, they obtained a randomized $(1-1/e-\varepsilon)$-approximation algorithm using $\widetilde{O}(n \cdot 2^{O(1/\varepsilon^4)})$ value oracle queries and $\TO(n + r^{3/2})$ independence oracle queries, which is nearly-linear in $n$ when $r = O(n^{2/3})$.
We note that, unlike the algorithms of~\cite{badanidiyuru2014fast, buchbinder2017comparing, kobayashi_et_al:LIPIcs.ICALP.2024.100}, the value query complexities of these algorithms depend exponentially on $1/\varepsilon$.

Buchbinder and Feldman~\cite{buchbinder2025deterministic} posed, as an open question, whether the query complexity can be further improved to be nearly-linear in $n$ for all values of $r$, rather than only for $r = \tilde{O}(n^{2/3})$ as in~\cite{buchbinder2025extending}.

\subsection{Our Results}
Table~\ref{tab:known-results} summarizes the best known complexities of $(1-1/e-\eps)$-approximation algorithms for monotone submodular maximization subject to a matroid constraint, together with our new results.
We develop a new local search algorithm, which we call \emph{protected local search}, based on the potential function of Buchbinder and Feldman~\cite{buchbinder2025deterministic}.
Our main result is the following theorem.

\begin{theorem} \label{thm:main_r}
    For any $\eps > 0$, there is a randomized algorithm that achieves a $(1-1/e-\eps)$-approximation for maximizing a monotone submodular function subject to a matroid constraint and uses $O(n \cdot \Poly(1/\eps, \log n))$ value and independence oracle queries.
\end{theorem}

This is the first $(1-1/e-\varepsilon)$-approximation algorithm with both independence-oracle and value-oracle query complexities nearly-linear in $n$ and polynomial in $1/\varepsilon$.
In particular, this resolves the open question posed by Buchbinder and Feldman~\cite{buchbinder2025deterministic}.

Our algorithm can also be derandomized at the cost of an exponential dependence on $1/\varepsilon$ in the value-oracle query complexity.

\begin{theorem} \label{thm:main_d}
    For any $\eps > 0$, there is a deterministic algorithm that achieves a $(1-1/e-\eps)$-approximation for maximizing a monotone submodular function subject to a matroid constraint and uses $O(n \cdot \Poly(\log n) \cdot 2^{O(1/\eps)})$ value-oracle queries and $O(n \cdot \Poly(1/\eps, \log n))$ independence-oracle queries.
\end{theorem}

\begin{table}[t]
\centering
\small
\setlength{\tabcolsep}{4pt}
\renewcommand{\arraystretch}{1.22}
\begin{tabularx}{\linewidth}{@{}Xcccc@{}}
\toprule
\textbf{Reference}  & \textbf{Value queries} & \textbf{Independence queries} & \textbf{Type} \\
\midrule
Calinescu, Chekuri, P\'{a}l and Vondr\'{a}k~\cite{calinescu2011maximizing}
 & $\operatorname{poly}(n)$ & $\operatorname{poly}(n)$ & R \\
Filmus and Ward~\cite{filmus2014monotone}
 & $\operatorname{poly}(n)$ & $\operatorname{poly}(n)$ & R \\
Badanidiyuru and Vondr\'{a}k~\cite{badanidiyuru2014fast}
& $\widetilde O_\varepsilon(nr)$
  & $\widetilde O_\varepsilon(n+r^2)$ & R \\
Buchbinder, Feldman, and Schwartz~\cite{buchbinder2017comparing}
 & $\widetilde O_\varepsilon(n\sqrt r)$
  & $\widetilde O_\varepsilon(n\sqrt r+r^2)$ & R \\
Kobayashi and Terao~\cite{kobayashi_et_al:LIPIcs.ICALP.2024.100}
& $\widetilde O_\varepsilon(n\sqrt r)$
  & $\widetilde O_\varepsilon(n\sqrt r)$ & R \\
Buchbinder and Feldman~\cite{buchbinder2025deterministic}
 & $\widetilde O(nr \cdot 2^{O(1/\eps)})$
  & $\widetilde O_\varepsilon(nr)$ & D \\
Buchbinder and Feldman~\cite{buchbinder2025deterministic}
 & $\widetilde O((n+r\sqrt n) \cdot 2^{O(1/\eps)})$
  & $\widetilde O_\varepsilon(n+r\sqrt n)$ & R \\
Buchbinder and Feldman~\cite{buchbinder2025extending}
& $\widetilde O(n \cdot 2^{O(1/\eps^4)})$
  & $\widetilde O_\varepsilon(n+r^{3/2})$ & R \\
Theorem~{\ref{thm:main_d}}
 & $\widetilde O(n \cdot 2^{O(1/\eps)})$
& $\widetilde O_\varepsilon(n)$ & D \\
Theorem~{\ref{thm:main_r}}
& $\widetilde O_\varepsilon(n)$
& $\widetilde O_\varepsilon(n)$ & R \\
\bottomrule
\end{tabularx}
\caption{Oracle-query complexities for maximizing a nonnegative monotone submodular function subject to a general matroid constraint.
Here, D and R denote deterministic and randomized algorithms, respectively.
Note that the $\tilde{O}$ notation hides polylogarithmic factors in $n$, and the $\tilde{O}_{\varepsilon}$ notation additionally hides polynomial factors in $\varepsilon^{-1}$.}
\label{tab:known-results}
\vspace{3pt}
\begin{minipage}{\linewidth}
\end{minipage}
\end{table}

\subsection{Additional Related Work}
Buchbinder and Feldman~\cite{buchbinder2025extending} also gave a deterministic algorithm that computes a value $V$ with $(1-1/e-\eps)f(\mathrm{OPT}) \le V \le f(\mathrm{OPT})$ using $\tO(n \cdot 2^{O(1/\eps^4)})$ value-oracle and $\TO(n)$ independence-oracle queries. Unlike our algorithm, however, it only estimates the optimal value and does not output a feasible solution.

Recently, faster algorithms based on the continuous greedy framework have been developed for several important classes of matroids.
Specifically, $(1 - 1/e - \varepsilon)$-approximation algorithms using nearly-linear value-oracle queries and arithmetic operations are known for graphic and partition matroids~\cite{ene2018towards} and for laminar and transversal matroids~\cite{henzinger2023faster}.

Ganz Rozenman, Kulik, Schwartz, and Singh~\cite{ganz2026poisson} proposed a different approach based on a stochastic Poisson process, which requires neither discretization nor rounding.
For partition matroids, their approach yields a $(1-1/e-\varepsilon)$-approximation algorithm using $O(n \cdot \Poly(1/\eps))$ value oracle queries, which is linear in $n$.

\section{Preliminaries}

\subparagraph*{Basic Notation.}
Let $\mathbb{R}_{+}$ denote the set of non-negative real numbers.
Throughout this paper, let $\N$ be a finite set and let $n$ denote its cardinality.
For a set $A \subseteq \N$ and an element $v \in \N$, we will often write $A + v := A \cup \{ v \}$ and $A - v := A \setminus \{ v \}$.
For a positive integer $k$, let $[k] := \{1, \dots, k\}$.
For a set function $h$, a set $S$, and an element $v$, we write $h(v \mid S) := h(S + v) - h(S)$.
Note that $h(v \mid S) = 0$ if $v \in S$.

We assume without loss of generality that $0 < \eps \le 1/4$.
We say that an event happens \emph{with high probability} (w.h.p.) if it happens with probability at least $1-(\eps/n)^{c}$, where $c>0$ is an arbitrarily large constant that affects only the constant factors in the query complexity.

\subparagraph*{Submodular Functions.}
Let $f \colon 2^\N \rightarrow \mathbb{R}_{+}$ be a set function on a finite ground set $\N$ of size $n$.
The function is \emph{submodular} if $f(A) + f(B) \geq f(A \cup B) + f(A \cap B)$ for any two subsets $A, B \subseteq \N$.
The function is \emph{monotone} if $f(A) \leq f(B)$ for any subsets $A \subseteq B \subseteq \N$.
In this paper, we only consider monotone submodular functions.
For such a function, $f(v \mid A) \ge f(v \mid B)$ holds for any $A \subseteq B \subseteq \N$ and $v \in \N$; we refer to this inequality simply as submodularity.

\subparagraph*{Matroids.}
A pair $\M = (\N, \mathcal{I})$ of a finite set $\N$ and a non-empty set family $\mathcal{I} \subseteq 2^{\N}$ is called a {\em matroid} if the following properties are satisfied.

\begin{description}
\item[(Downward closure property)]
if $S \in \mathcal{I}$ and $S' \subseteq S$, then $S' \in \mathcal{I}$.

\item[(Augmentation property)]
if $S, S' \in \mathcal{I}$ and $|S'| < |S|$, then there exists $v \in S \setminus S'$ such that $S' + v \in \mathcal{I}$.
\end{description}

A set $S \subseteq \N$ is called {\em independent} if $S \in \mathcal{I}$ and {\em dependent} otherwise.
The {\em rank} of $\M$ is defined as the size of a largest independent set.
In addition, for a subset $S \subseteq \N$, the rank of $S$ is defined as the size of a largest independent set contained in $S$.
Inclusion-wise maximal independent sets are called {\em bases}.
Note that every basis has the same size.
Throughout this paper, $r$ denotes the rank of $\M$.
Before running the algorithm, we remove all loops of $\M$, i.e., elements $u\in\N$ with $\{u\}\notin\I$.
Hence, we assume without loss of generality that $\M$ has no loops.

\subparagraph*{Problem.}
We consider the problem of maximizing a monotone submodular function $f \colon 2^\N \to \mathbb{R}_+$ subject to a matroid $\M = (\N, \I)$, that is, $\max\{ f(S) \mid S \in \I \}$.
We denote by $\OPT \in \I$ an optimal solution.

\subparagraph*{Oracles.}
When we consider the submodular maximization problem, we assume that the submodular function $f$ is given as a value oracle, which takes as input any subset $S \subseteq \N$ and outputs $f(S)$.
We also assume that we access a matroid $\M$ through an oracle.
Given a subset $S \subseteq \N$, an {\em independence oracle} outputs whether $S \in \I$ or not.


\subsection{Basic Facts on Submodular Functions}

We use the following basic inequalities for submodular functions.

\begin{fact}[From {\cite[Lemmas~2.2 and 2.3 in the journal version]{buchbinder2025deterministic}}] \label{fact:submod}
Let $f\colon 2^\N \to \mathbb{R}_{+}$ be a submodular function.
\begin{enumerate}[(a)]
\item If $S\subseteq T\subseteq \N$, then $\sum_{u\in T\setminus S}f(u\mid T-u)\le f(T)-f(S)$.
\item For all $S,T\subseteq \N$, $\sum_{u\in T}f(u\mid S)\ge f(S\cup T)-f(S)$.
\end{enumerate}
\end{fact}

\subsection{Binary Search Technique} \label{subsec:bs}

Nguyen~\cite{nguyen2019note} and Chakrabarty, Lee, Sidford, Singla, and Wong~\cite{chakrabarty2019faster} independently presented the so-called {\em binary search technique}, which is useful for improving the query complexity of matroid problems.

\begin{lemma}[Binary Search Technique {\cite{nguyen2019note, chakrabarty2019faster}}] \label{lem:bstech}
    Let $\M = (\N, \I)$ be a matroid.
    Let $S \in \I, T \subseteq S$, and $v \in \N \setminus S$.
    If there exists an element $u\in T$ such that $S+v-u\in\I$, then such an element can be found using $O(\log |T|)$ independence oracle queries.
\end{lemma}

The idea behind Lemma~\ref{lem:bstech} is to perform binary search on $T$.
We skip the proof in this paper.

\subsection{The Lifted Matroid and Potential Function of Buchbinder and Feldman} \label{sec:potential}
In this subsection, we describe the lifted matroid and the potential function, which were used in the non-oblivious local search algorithm of Buchbinder and Feldman~\cite{buchbinder2025deterministic}, and present some of their useful properties.
Our algorithm also performs local search on this lifted matroid, guided by the same potential function, but it updates the solution in an entirely different way, as we explain in Section~\ref{sec:technical_overview}.

Throughout the rest of this paper, we set $\ell = \lceil 1/\eps \rceil$, so that $1/\ell \le \eps$ and $\ell = O(1/\eps)$.
We use the following coefficients of Buchbinder and Feldman~\cite{buchbinder2025deterministic}:
\[
\alpha_0=0,\qquad\alpha_k=\frac{(1+1/\ell)^{k-1}}{\binom{\ell-1}{k-1}}\quad(k\in[\ell]).
\]
We also write $\kappa = (\ell+1)\alpha_\ell = \ell(1+1/\ell)^{\ell}$.
Here, we have $1 - \ell/\kappa = 1-(1+1/\ell)^{-\ell} \ge 1 - 1/e - \eps$ (see Fact~\ref{fact:coef}).

\subparagraph*{Lifted Matroid.}
We also use the lifted matroid construction of Buchbinder and Feldman~\cite{buchbinder2025deterministic}.
Let $\NL=\N\times[\ell]$. 
The element $(u,i)$ is called the \emph{copy} of $u$ with color $i$. We use $u,w$ for elements of $\N$ and $x,y$ for elements of $\NL$. For $J\subseteq[\ell]$ and $S\subseteq\NL$, let
\[
\pi_J(S)=\{u\in\N \mid (u,i)\in S\text{ for some }i\in J\} \quad \text{and} \quad \pi(S)=\pi_{[\ell]}(S).
\]
Let $\ML=(\NL,\IL)$, where $S\in\IL$ if and only if $S$ contains at most one copy of each element of $\N$ and $\pi(S)\in\I$.
An independence oracle query to $\ML$ can be answered using at most one independence oracle query to $\M$, since a set containing two copies of the same element does not belong to $\IL$.
Note that $|\pi(S)| = |S|$ for every $S \in \IL$, and that $S \times \{j\} \in \IL$ for every $S \in \I$ and $j \in [\ell]$.
The following lemma guarantees that the lifting preserves the matroid structure.

\begin{lemma}[From {\cite[Lemma~3.5 in the journal version]{buchbinder2025deterministic}}] \label{lem:lifted}
    $\ML$ is a matroid, and it has the same rank $r$ as $\M$.
\end{lemma}

\textbf{Our algorithm runs on the lifted matroid $\ML$ instead of $\M$}: it maintains a basis $B$ of $\ML$ and finally outputs $\pi(B)$, which is a basis of $\M$ since $\pi(B)\in\I$ and $|\pi(B)|=|B|=r$.

\subparagraph*{Potential Function.}
We also use the potential function $g\colon 2^{\NL} \to \mathbb{R}$ of Buchbinder and Feldman~\cite{buchbinder2025deterministic},\footnote{This function is denoted by $g'$ in~\cite{buchbinder2025deterministic}.} which is defined for every $S \subseteq \NL$ as
\[
g(S)=\sum_{J\subseteq[\ell]}\alpha_{|J|} \cdot f\bigl(\pi_J(S)\bigr).
\]
The following property of $g$ was proved.

\begin{lemma}[From {\cite[Lemma~3.4 in the journal version]{buchbinder2025deterministic}}] \label{lem:g}
    The function $g$ is non-negative, monotone, and submodular.
\end{lemma}

In the analysis, we also use the auxiliary function $z\colon 2^{\NL} \to \mathbb{R}$, defined by
\[
z(S)=\sum_{J\subseteq[\ell]}\frac{|J|}{\ell}\,\alpha_{|J|} \cdot f\bigl(\pi_J(S)\bigr).
\]
Since $0 \le \frac{|J|}{\ell}\alpha_{|J|} \le \alpha_{|J|}$ for every $J \subseteq [\ell]$ and each function $S \to f(\pi_J(S))$ is monotone, we have
\begin{equation}\label{eq:z-vs-g}
0\le z(X+y)-z(X)\le g(X+y)-g(X) \qquad \text{for any $X\subseteq\NL$ and $y\in\NL$}.
\end{equation}

We use the following properties of the coefficients $\alpha_k$. Here, recall that $\kappa = \ell(1+1/\ell)^{\ell}$.

\begin{restatable}{fact}{coef}\label{fact:coef}
\leavevmode
\begin{enumerate}[(a)]
\item $(\ell-k)\alpha_{k+1}=k(1+1/\ell)\alpha_k$ for $1\le k<\ell$.
\item For every $i\in[\ell]$, $\sum_{J\ni i}\alpha_{|J|}=\kappa-\ell$. We also have $\sum_{J}\frac{|J|}{\ell}\alpha_{|J|}=\kappa-\ell$.
\item $\sum_{J}\alpha_{|J|}=O(\ell^2)$.
\item $\ell\le\kappa\le e\ell$ and $\ell/\kappa\le1/e+1/\ell$. In particular, $1-\ell/\kappa\ge1-1/e-\eps$.
\end{enumerate}
\end{restatable}





We defer the proof of \Cref{fact:coef} to Appendix~\ref{appendix:potential}.
Next, we describe two basic properties of the potential function $g$.

\begin{restatable}{observation}{marginal}\label{obs:marginal}
For $x=(u,i)\in\NL$ and $X\subseteq\NL$, $\ds g(x\mid X-x)=\sum_{J\ni i}\alpha_{|J|}\cdot f\bigl(u\mid\pi_J(X-x)\bigr)$.
\end{restatable}


\begin{restatable}{observation}{range}\label{obs:range}
For every $X\subseteq\NL$, we have $g(X)\le O(\ell^2)\cdot f(\pi(X))$. For every $y=(u,i)\in\NL$ and $X\subseteq\NL$, we have $g(y\mid X)\le(\kappa-\ell)\cdot f(\{u\})$.
\end{restatable}

We defer the proofs of Observations \ref{obs:marginal} and \ref{obs:range} to Appendix~\ref{appendix:potential}.


\subsection{Comparison Inequality} \label{subsec:comp}

In this subsection, we derive a bound on the potential function~$g$, which is the key inequality in the analyses of both the algorithm of Buchbinder and Feldman~\cite{buchbinder2025deterministic} and our algorithm. 
We remark that a special case of \Cref{cor:compare} also plays an important role in their analysis (see \Cref{sec:technical_overview}).

\begin{restatable}{lemma}{deladd}\label{lem:del-add}
\leavevmode
\begin{enumerate}[(a)]
\item For every $C\in\IL$, $\sum_{x\in C}g(x\mid C-x)+z(C)\le\kappa\cdot f\bigl(\pi(C)\bigr)$.
\item For every $H\subseteq\NL$, $\ds \frac1\ell\sum_{j=1}^{\ell}\sum_{u\in\OPT}g\bigl((u,j)\bigm| H-(u,j)\bigr)\ge(\kappa-\ell)\cdot f(\OPT)-z(H).$
\end{enumerate}
\end{restatable}

We defer the proof of \Cref{lem:del-add} to Appendix~\ref{appendix:deladd}.
Adding the two inequalities of Lemma~\ref{lem:del-add} yields the following.

\begin{corollary}\label{cor:compare}
For every $C\in\IL$ and every $H\subseteq\NL$,
\[
(\kappa-\ell)\cdot f(\OPT)\le\kappa\cdot f\bigl(\pi(C)\bigr)+D_1+D_2,
\]
where $D_1=z(H)-z(C)$ and $\ds D_2=\frac1\ell\sum_{j=1}^{\ell}\sum_{u\in\OPT}g\bigl((u,j)\bigm| H-(u,j)\bigr)-\sum_{x\in C}g(x\mid C-x)$.
\end{corollary}

\subsection{Estimating {$g(x\mid X-x)$}}\label{sec:estimation}

Our algorithm needs to evaluate $g(x\mid X-x)$.
By Observation~\ref{obs:marginal}, $g(x\mid X-x)$ is a sum of $2^{\ell-1}$ terms, each of which can be computed using two value-oracle queries; hence, $g(x\mid X-x)$ can be computed exactly using $2^{\ell}=2^{O(1/\eps)}$ value-oracle queries (see Lemma~\ref{lem:exact}).
Our deterministic algorithm (\Cref{thm:main_d}) uses this exact computation.
To improve the dependence on $1/\eps$ from exponential to polynomial, our randomized algorithm instead estimates $g(x\mid X-x)$ by sampling (see Lemma~\ref{lem:estimate}).

\begin{lemma}[Exact computation]\label{lem:exact}
For any $x\in\NL$ and $X\subseteq\NL$, the value $g(x\mid X-x)$ can be computed deterministically using $2^\ell$ value oracle queries.
\end{lemma}

\begin{proof}
Let $x=(u,i)$. By Observation~\ref{obs:marginal}, $g(x\mid X-x)$ is determined by the values $f(\pi_J(X-x)+u)$ and $f(\pi_J(X-x))$ for the $2^{\ell-1}$ sets $J\subseteq[\ell]$ that contain $i$.
\end{proof}

\begin{restatable}{lemma}{est} \label{lem:estimate}
There is a randomized algorithm $\Est(x,X)$ which, given $x\in\NL$ and $X\subseteq\NL$, uses $O(\eps^{-3}\log(n/\eps))$ value oracle queries and w.h.p.\ outputs $\widehat g$ such that $\frac{g(x\mid X-x)}{1+\eps}\le\widehat g\le g(x\mid X-x)$.
Since each call uses fresh randomness, this holds even if $(x,X)$ is chosen depending on the outcomes of previous calls.
\end{restatable}

Roughly speaking, the procedure $\Est$ draws $O(\ell\eps^{-2}\log(n/\eps))$ independent random samples, computes a value for each sample using value oracle queries, and outputs the average of the computed values.
We defer the proof of \Cref{lem:estimate} to Appendix~\ref{appendix:estimate}.

We call an estimate \emph{good} if it satisfies the inequalities in Lemma~\ref{lem:estimate}.

\section{Technical Overview: Comparison with the Local Search of Buchbinder and Feldman~{\cite{buchbinder2025deterministic}}} \label{sec:technical_overview}

In this section, we first explain the analysis of the local search of Buchbinder and Feldman in terms of \Cref{cor:compare}, and then outline how our algorithm uses the same corollary in its analysis.
\Cref{cor:compare} states that, for every $C\in\IL$ and every $H\subseteq\NL$, 
\[
(\kappa-\ell)\cdot f(\OPT)\le\kappa\cdot f\bigl(\pi(C)\bigr)+D_1+D_2,
\]
where $D_1=z(H)-z(C)$ and $\displaystyle D_2=\frac1\ell\sum_{j=1}^{\ell}\sum_{u\in\OPT}g\bigl((u,j)\bigm| H-(u,j)\bigr)-\sum_{x\in C}g(x\mid C-x)$.

\subparagraph*{The Analysis of Local Search of Buchbinder and Feldman~\cite{buchbinder2025deterministic}.}
Roughly speaking, they compute a set $S\in\IL$ by local search, and their analysis essentially applies the inequality in \Cref{cor:compare} with $C=H=S$.
In this case, we have $D_1=0$.
Moreover, the inequality satisfied by the set $S$ obtained by local search~\cite[Inequality~(3.4)]{buchbinder2025deterministic} implies that $D_2\le\eps\ell\cdot f(\OPT)$.\footnote{To see this, note that Inequality~(3.4) of~\cite{buchbinder2025deterministic} states that $\sum_{v \in T} g(v \mid S - v) - \sum_{u \in S} g(u \mid S - u) \le \eps' \cdot g(\OPT')$ for every $T \in \IL$, where $\eps' = \eps / (e(1 + \ln \ell))$ and $\OPT'$ is a set in $\IL$ maximizing $g$. Applying this with $T = \OPT \times \{j\} \in \IL$ for each $j \in [\ell]$ and averaging over $j$, we obtain $D_2 \le \eps' \cdot g(\OPT')$. Combining this with $g(\OPT') \le e\ell(1 + \ln \ell) \cdot f(\OPT)$, shown at the end of the proof of~\cite[Lemma~3.6]{buchbinder2025deterministic}, yields the claimed bound $D_2 \le \eps\ell \cdot f(\OPT)$.}
Combining these bounds and dividing by $\kappa\ge\ell$, we obtain $f(\pi(S))\ge(1-\ell/\kappa-\eps)\cdot f(\OPT)$.
Since $1-\ell/\kappa\ge1-1/e-\eps$ by \Cref{fact:coef}(d), this yields a $(1-1/e-O(\eps))$-approximation.

However, their local search for finding a set $S$ satisfying~\cite[Inequality~(3.4)]{buchbinder2025deterministic} uses $\tilde{O}(nr \cdot 2^{O(1/\eps)})$ queries, or $\tilde{O}((n+r\sqrt{n}) \cdot 2^{O(1/\eps)})$ queries with randomization, which is not nearly-linear in $n$ when $r$ is large.

\subparagraph*{The Analysis of Our Protected Local Search.}
Our \emph{protected local search} performs $T=O(\varepsilon^{-3})$ phases, each of which is a single pass over decreasing thresholds $\Theta$ in which every element $y$ is processed at most once: $y$ is swapped in for an element $x$ of the current basis only if removing $x$ costs less than about $\Theta$, and otherwise $x$ is \emph{protected}, i.e., kept in the basis until the end of the phase; see \Cref{alg:phase} for details.

Fix a single phase, and let $B_0$ and $B_1$ be the bases at the beginning and the end of this phase, respectively. Let $\Delta = g(B_1) - g(B_0)$ be the increase in $g$ during this phase.
We show that $\Delta\ge0$ and $f\bigl(\pi(B_0)\bigr)\ge\Bigl(1-\frac1e-O(\eps)\Bigr)\cdot f(\OPT)-O(\Delta)$ (\Cref{lem:phase}).
To prove the second inequality, we apply \Cref{cor:compare} to the sets $C$ and $H$ at the end of the phase.
Since $C\subseteq B_0$ and $f$ is monotone, we have $f(\pi(B_0))\ge f(\pi(C))$, and hence it suffices to upper-bound $D_1 + D_2$.
For the upper bound on $D_1+D_2$, see the paragraph \emph{Plan of the Proof of \Cref{lem:phase}} in \Cref{sec:phase}.

\Cref{lem:phase} means that if $\Delta\le\eps\cdot f(\OPT)$, then $\pi(B_0)$ is a $(1-1/e-O(\eps))$-approximate solution; otherwise, the phase increases $g$ by more than $\eps\cdot f(\OPT)$.
Since $\Delta=g(B_1)-g(B_0)\ge0$ for every phase and $g(B)\le O(\ell^2)\cdot f(\OPT)$ for every basis $B$ of $\ML$ (by Observation~\ref{obs:range}), the total increase of $g$ over $T$ phases is at most $O(\ell^2)\cdot f(\OPT)$.
Hence, for $T=\ell^3$, some phase satisfies $\Delta\le O(\ell^2/T)\cdot f(\OPT)=O(\eps)\cdot f(\OPT)$ since $\ell\ge1/\eps$, and its starting basis $B_0$ yields a $(1-1/e-O(\eps))$-approximation.
Our algorithm returns the best among the starting bases of the $T$ phases.

\section{Our Protected Local Search Algorithm}\label{sec:algorithm}

\subsection{Parameters}\label{sec:params}

We first construct a basis $\Binit$ of $\M$ greedily using $n$ independence queries, and compute $\fmax=\max_{u \in \N}f(\{u\})$ using $n$ value queries.
Since we assume that $\M$ has no loops, we have $\fmax\le f(\OPT)$.
If $\fmax=0$, then $0\le f(\emptyset)\le f(\{u\})=0$ by monotonicity, and Fact~\ref{fact:submod}(b) gives $f(X)\le\sum_{u\in X}f(\{u\})=0$ for every $X\subseteq\N$; hence we simply return $\Binit$. 

\begin{table}[t]
\centering
\caption{Parameters of the algorithm.}\label{tab:params}
\begin{tabular}{@{}lll@{}}
\toprule
Symbol & Value & Role\\
\midrule
$\ell$ & $\lceil1/\eps\rceil$ & number of colors\\
$\Tmax$ & $(\kappa-\ell)\cdot\fmax$ & upper limit for thresholds\\
$\Tmin$ & $\fmax/r$ & lower limit for thresholds\\
$\Theta_t$ & $\Tmax\cdot (1+\eps)^{-(t-1)}$, $t\in[\tmax]$ & thresholds, in decreasing order\\
$T$ & $\ell^3$ & number of phases\\
\bottomrule
\end{tabular}
\end{table}

The parameters in our algorithm are listed in Table~\ref{tab:params}.
The thresholds are $\Theta_t=\Tmax(1+\eps)^{-(t-1)}$ for $t\in[\tmax]$, where $\tmax$ is the largest $t$ with $\Theta_t\ge\Tmin$.
Note that $\Theta_{\tmax}<(1+\eps)\Tmin$.
Since $\Tmax/\Tmin\le\kappa r\le e\ell r$ by \Cref{fact:coef}(d) and $\ln(1+\eps)\ge\eps/2$ for $\eps<1$, the number of thresholds is $\ds \tmax=O\Bigl(\frac{\log(\ell r)}{\eps}\Bigr)=O\Bigl(\frac{\log(n/\eps)}{\eps}\Bigr)$.
By Observation~\ref{obs:range}, we have $g(y\mid X)\le (\kappa - \ell) \cdot \fmax = \Tmax$ for every $y\in\NL$ and $X\subseteq\NL$.

\subsection{One Phase}\label{sec:one-phase}

Our algorithm consists of $T$ phases, each of which is described in \Cref{alg:phase}.

\begin{algorithm}[t]
\caption{$\Phase(B_0)$}\label{alg:phase}
\begin{algorithmic}[1]
\Require a basis $B_0$ of $\ML$
\State $A\gets\emptyset$;\quad $C\gets B_0$;\quad mark every $y\in\NL$ as unprocessed
\Statex \emph{\quad Throughout: $B=A\cup C$, \ $H=B_0\cup A$, \ $\gamma(y)=g(y\mid H-y)$, \ $\lambda(x)=g(x\mid C-x)$.}
\For{$\Theta=\Theta_1,\Theta_2,\dots,\Theta_{\tmax}$}
  \For{each unprocessed $y\in\NL$, in arbitrary order}
    \If{$\hg(y)<\Theta$} \label{line:skip} \algorithmiccontinue \Comment{$y$ stays unprocessed}
    \EndIf
    \While{$y\notin A$ and $A+y\in\IL$} \Comment{processing of $y$}
      \If{$y\in B$}
        \State $x\gets y$
      \Else
        \State $x\gets$ an element of $C\setminus A$ such that $B-x+y\in\IL$ \label{line:partner} \Comment{by the binary search technique (Lemma~\ref{lem:bstech})}
      \EndIf
      \If{$\hla(x)>\Theta/(1+\eps)^2$} \label{line:test}
        \State $A\gets A+x$ \label{line:protect} \Comment{protect $x$}
      \Else
        \State $C\gets C-x$;\quad $A\gets A+y$ \label{line:insert} \Comment{exchange $x$ for $y$}
        \State \algorithmicbreak
      \EndIf
    \EndWhile
    \State mark $y$ as processed
  \EndFor
\EndFor
\State \Return $B_1=A\cup C$
\end{algorithmic}
\end{algorithm}

A phase starts from a basis $B_0$ of $\ML$ and maintains the following sets.
\begin{description}
\item[$A$.] The elements that stay in the current basis until the end of the phase; $A$ never shrinks.
An element enters $A$ in one of two ways: it is \emph{protected} if it is added to $A$ in Line~\ref{line:protect}, and it is \emph{inserted} if it is added to $A$ in Line~\ref{line:insert}.
\item[$C\subseteq B_0$.] The elements of $B_0$ that have not been deleted. $C$ never grows.
\item[$B=A\cup C$.] The current basis.
\item[$H=B_0\cup A$.] The set on which gains are evaluated.
\end{description}
The sets $A$ and $C$ may intersect, since a protected element belongs to both. For the current sets, we write
\[
\gamma(y)=g(y\mid H-y)\quad\text{(gain of $y$)}\qquad\text{and}\qquad\lambda(x)=g(x\mid C-x)\quad\text{(loss of $x$)}.
\]
We call $\gamma(y)$ the gain of $y$ also when $y\in H$, and $\lambda(x)$ the loss of $x$ also when $x\notin C$. 
We write $\hg(y)$ and $\hla(x)$ for the estimates returned by $\Est(y,H)$ and $\Est(x,C)$, respectively. 

In words, for each threshold, the phase scans the elements that are still unprocessed. 
An element $y \in \NL$ whose estimated gain $\hg(y)$ is below the threshold $\Theta$ is skipped, and it is scanned again at the next threshold. 
Otherwise, the element $y$ is \emph{processed}: as long as $y \notin A$ and $A+y$ is independent, the phase sets $x=y$ if $y\in B$, and otherwise picks an element $x\in C\setminus A$ such that $B-x+y\in\IL$.
The phase protects $x$ if the estimated loss $\hla(x)$ exceeds $\Theta/(1+\eps)^2$, and otherwise it exchanges $x$ for $y$ and stops. 
Afterwards, $y$ is marked as processed and is not examined again in this phase.

The following lemma plays a crucial role in the analysis of our algorithm.

\begin{observation}[Monotonicity of {$\gamma(y)$} and {$\lambda(x)$}]\label{obs:time}
During a phase, $H$ only grows and $C$ only shrinks. Hence, by submodularity of $g$ (Lemma~\ref{lem:g}), the gain $\gamma(y)$ of each $y$ can only decrease, and the loss $\lambda(x)$ of each $x$ can only increase.
\end{observation}

\subsection{The Whole Algorithm}

The whole algorithm is described in \Cref{alg:main}.
It first constructs a basis $\Binit$ of $\M$, computes $\fmax$, and sets the parameters as in Table~\ref{tab:params}.
It then assigns color $1$ to every element of $\Binit$, which yields a basis $B^{(0)}=\Binit\times\{1\}$ of $\ML$, and runs $T$ phases starting from $B^{(0)}$, where each phase takes the basis produced by the previous phase as its input.
Finally, it returns the best projected basis $\pi(B^{(t)})$ with respect to $f$.

\begin{algorithm}[t]
\caption{Protected local search}\label{alg:main}
\begin{algorithmic}[1]
\State Construct a basis $\Binit$ of $\M$; compute $\fmax$
\If{$\fmax=0$} \Return $\Binit$
\EndIf
\State set the parameters as in Table~\ref{tab:params}
\State $B^{(0)}\gets\Binit\times\{1\}$ \Comment{every element of the initial basis gets color $1$}
\For{$t=0,1,\dots,T-1$}
  \State $B^{(t+1)}\gets\Phase(B^{(t)})$
\EndFor
\State \Return $\pi\bigl(B^{(t^*)}\bigr)$, where $t^*\in\arg\max_{0\le t<T}f\bigl(\pi(B^{(t)})\bigr)$
\end{algorithmic}
\end{algorithm}

\subsection{Invariants and Query Complexity of One Phase}

In this subsection, we show some invariants maintained during a phase and then bound the number of oracle queries used in a phase.
Both the invariants and the bound hold regardless of whether the estimates are good.

\begin{lemma}[Invariants]\label{lem:invariants}
Throughout a phase, $A\in\IL$, $C\subseteq B_0$, $B=A\cup C$ is a basis of $\ML$, and $C\subseteq B\subseteq H$. Moreover, $A$ and $H$ only grows and $C$ only shrinks, and each iteration of the while loop adds a new element to $A$. In particular, the body of the while loop is executed at most $r$ times in total during a phase.
\end{lemma}

\begin{proof}
The sets $A$ and $C$ change only in iterations of the while loop, and the invariants hold at the beginning of the phase, where $A=\emptyset$ and $B=C=H=B_0$. We show that each iteration maintains them. Consider an iteration processing $y$; by the loop condition, $y\notin A$ and $A+y\in\IL$.

We first show that the element $x$ is well-defined and satisfies $x\in C\setminus A$.
If $y\in B$, then $x=y\in C\setminus A$, because $B=A\cup C$ and $y\notin A$. Now suppose that $y \notin B$. It remains to show that an element $x$ as in Line~\ref{line:partner} always exists.
To prove this, we use the augmentation property of the matroid $\ML$: by repeatedly adding elements of $B$ to $A+y$, we obtain a basis $B'$ of $\ML$ with $A+y\subseteq B'\subseteq B+y$. Since $|B'|=|B|$ and $y\in B'$, we have $B'=B-x'+y$ for some $x'\in B$, and $x'\notin A$ because $A\subseteq B'$. Hence $x'\in B\setminus A=C\setminus A$ satisfies $B-x'+y\in\IL$, and thus an element $x$ as in Line~\ref{line:partner} always exists.

We next consider the update.
\begin{itemize}
\item If $x$ is protected (Line~\ref{line:protect}), then $A$ becomes $A+x\subseteq B$, which is independent. Since $x\in C\subseteq B_0$, none of $B=A\cup C$, $C$, and $H=B_0 \cup A$ changes.
\item If $x\neq y$ is exchanged for $y$ (Line~\ref{line:insert}), then $y\notin B$, and $B$ becomes $(A+y)\cup(C-x)=B-x+y$. This is a basis of $\ML$ that contains the new set $A+y$. The set $H$ becomes $H+y$, and $C$ becomes $C-x$.
\item If $x=y$ is exchanged for $y$, then $y$ moves from $C$ to $A$, so $B$ and $H$ do not change, and $A+y\subseteq B$ is independent.
\end{itemize}
In all cases, $C\subseteq B_0$ and $C\subseteq B\subseteq A\cup B_0=H$ still hold, $A$ and $H$ do not shrink, and $C$ does not grow. Finally, every iteration adds an element not in $A$ (namely, $x$ or $y$) to $A$. Since $A\in\IL$ has at most $r$ elements, there are at most $r$ iterations.
\end{proof}

\begin{lemma}[Query complexity of one phase]\label{lem:phase-queries}
$\Phase(B_0)$ makes $O(\eps^{-2}n\log(n/\eps))$ calls to $\Est$. Moreover, $\Phase(B_0)$ uses $O(\eps^{-5}n\log^2(n/\eps))$ value-oracle queries and $O(\eps^{-1}n+r\log r)$ independence-oracle queries.
\end{lemma}

\begin{proof}
We first count the value-oracle queries, which are used only in the calls to $\Est$. For each of the $\tmax$ thresholds, the gain of each unprocessed element is estimated once, so the number of gain estimates is at most $|\NL|\cdot\tmax\le n\ell\tmax=O(\eps^{-2}n\log(n/\eps))$. A loss is estimated once per iteration of the while loop, so by Lemma~\ref{lem:invariants} the number of loss estimates is at most $r\le n$. Hence, $\Phase(B_0)$ calls $\Est$ $O(\eps^{-2}n\log(n/\eps))$ times. By Lemma~\ref{lem:estimate}, each call uses $O(\eps^{-3}\log(n/\eps))$ value-oracle queries, and thus the number of value-oracle queries is $O(\eps^{-5}n\log^2(n/\eps))$.

We next count the independence-oracle queries. Recall that each independence-oracle query to $\ML$ is answered by at most one independence-oracle query to $\M$. Independence-oracle queries are used only to check the loop condition $A+y\in\IL$ and to find $x$ in Line~\ref{line:partner}.
\begin{itemize}
\item Each check of whether $A + y \in \IL$ either leads to an iteration of the while loop, which happens at most $r$ times by Lemma~\ref{lem:invariants}, or ends the processing of $y$, which happens at most once for each $y$, i.e., at most $|\NL|\le n\ell$ times.
\item Line~\ref{line:partner} is executed at most once per iteration, i.e., at most $r$ times. We apply Lemma~\ref{lem:bstech} to $\ML$ with $S=B$, $T=C\setminus A$, and $v=y$. A desired element exists by the proof of Lemma~\ref{lem:invariants}, and $|T|\le|B|=r$, so each execution uses $O(\log r)$ queries.
\end{itemize}
Hence, the number of independence-oracle queries is $O(n\ell+r\log r)=O(\eps^{-1}n+r\log r)$.
\end{proof}

\section{Analysis of One Phase}\label{sec:phase}

Throughout this section, we consider a single phase, and we assume that every estimate made during it is good.

\begin{lemma}[Phase lemma]\label{lem:phase}
Let $B_0$ and $B_1$ be the starting and ending bases of a phase in which all estimates are good, and let $\Delta=g(B_1)-g(B_0)$. Then $\Delta\ge0$ and
\[
f\bigl(\pi(B_0)\bigr)\ge\Bigl(1-\frac1e-O(\eps)\Bigr)\cdot f(\OPT)-O(\Delta).
\]
\end{lemma}

\subparagraph*{Notation.}
For a processed element $y$, let $\Theta(y)$ be the threshold at which it is processed. Let $E$ be the set of elements inserted by exchanges (Line~\ref{line:insert}) during the phase, and let $P$ be the set of protected (Line~\ref{line:protect}) elements. For $y\in E$, let $x_y$ be the element deleted from $C$ in its exchange, and let
\[
a_y=\gamma(y)\qquad\text{and}\qquad b_y=\lambda(x_y),
\]
where $\gamma$ and $\lambda$ are evaluated immediately before that exchange. Unless stated otherwise, $A$, $C$, $H$, $\gamma$, and $\lambda$ refer to the end of the phase.

\subparagraph*{Plan of the Proof of \Cref{lem:phase}.}
We apply Corollary~\ref{cor:compare} to the sets $C$ and $H$ at the end of the phase.
Since $C\subseteq B_0$ and $f$ is monotone, we have $f(\pi(B_0))\ge f(\pi(C))$, and hence it suffices to upper-bound $D_1+D_2$.
We first collect basic properties of a phase: Claim~\ref{clm:window} bounds the gain of each element in terms of the thresholds, and Claim~\ref{clm:exchange} shows that each exchange increases $g$ by at least an $\eps$-fraction of the gain of the inserted element, which in particular implies $\Delta\ge0$.
Next, we upper-bound $D_1=z(H)-z(C)$ by Claim~\ref{clm:charge}: each exchange changes $z(H)-z(C)$ by at most $O(a_y)$ while increasing $g$ by at least $\Omega(\eps)\cdot a_y$ (Claim~\ref{clm:exchange}), and hence $D_1=O(\Delta/\eps)$.
Finally, we upper-bound $D_2$ by Claims~\ref{clm:fixed}, \ref{clm:order}, and~\ref{clm:opt}.
Claim~\ref{clm:order} is the heart of the analysis; it plays the role of the approximate local optimality~\cite[Inequality~(3.4)]{buchbinder2025deterministic} in the analysis of Buchbinder and Feldman~\cite{buchbinder2025deterministic}, and bounds the total gain of any independent set by the sum of the thresholds at which the elements of $A$ entered $A$.
These thresholds are bounded by the losses of the protected elements (Claim~\ref{clm:fixed}) and by $\Delta$ for the inserted elements (Claim~\ref{clm:charge}), which yields Claim~\ref{clm:opt}.

Using \Cref{obs:time}, we first show that the gain of each element is bounded in terms of the thresholds.

\begin{claim}\label{clm:window}
If $y$ is processed at threshold $\Theta$, then $\Theta\le\gamma(y)\le(1+O(\eps))\cdot\Theta$ throughout its processing. Moreover, every element $y$ that is unprocessed at the end of the phase satisfies $\gamma(y)\le(1+O(\eps))\cdot\Tmin$ at the end of the phase.
\end{claim}

\begin{proof}
While $y$ is processed, $H=B_0\cup A$ can change only at an exchange in Line~\ref{line:insert}, after which the processing of $y$ immediately ends.
Indeed, a protection adds an element of $C\subseteq B_0$ to $A$, and hence does not change $H$.
Hence, $\gamma(y)=g(y\mid H-y)$ does not change during the processing of $y$.
The lower bound follows from $\gamma(y)\ge\hg(y)\ge\Theta$, where the second inequality holds because $y$ was not skipped in Line~\ref{line:skip}.
For the upper bound, if $\Theta=\Tmax$, then it follows from $\gamma(y)\le\Tmax$ (see Section~\ref{sec:params}).
Otherwise, $y$ was skipped in Line~\ref{line:skip} at the previous threshold $(1+\eps)\Theta$, i.e., $\hg(y)<(1+\eps)\Theta$ held at that time.
Hence, at that time, $\gamma(y)\le(1+\eps)\hg(y)<(1+\eps)^2\Theta$, and by Observation~\ref{obs:time}, $\gamma(y)$ has not increased since then.
Therefore, in both cases, $\Theta\le\gamma(y)\le(1+O(\eps))\cdot\Theta$ holds throughout the processing of $y$.

Let $y$ be an element that is unprocessed at the end of the phase.
Then, $y$ was skipped in Line~\ref{line:skip} at the last threshold $\Theta_{\tmax}<(1+\eps)\Tmin$, i.e., $\hg(y)<\Theta_{\tmax}$ held at that time.
Hence, at that time, $\gamma(y)\le(1+\eps)\hg(y)<(1+\eps)^2\Tmin$, and by Observation~\ref{obs:time}, $\gamma(y)$ has not increased since then.
Therefore, $\gamma(y)\le(1+O(\eps))\cdot\Tmin$ at the end of the phase.
\end{proof}

For $y\in E$, let $\Delta_y$ be the increase of $g(B)$ caused by the exchange that inserts $y$ (Line~\ref{line:insert}), i.e., $\Delta_y=g(B-x_y+y)-g(B)$, where $B$ denotes the basis just before the exchange.
Recall that $\Delta=g(B_1)-g(B_0)$.

\begin{claim}\label{clm:exchange}
\leavevmode
\begin{enumerate}[(a)]
\item In each iteration of the while loop during the processing of $y$, if $y\in B$ at that time, then $x=y$ is chosen in Line~\ref{line:partner}, and $y$ is protected (Line~\ref{line:protect}) in that iteration.
Consequently, just before each exchange in Line~\ref{line:insert}, we have $y\notin B$ and $x\ne y$.
\item For every $y\in E$, we have $\Theta(y)\le a_y$ and $b_y\le a_y/(1+\eps)$.
Moreover,
\[
\Delta_y\ge a_y-b_y\ge\frac{\eps}{1+\eps}\cdot a_y>0.
\]
\item $\Delta=\sum_{y\in E}\Delta_y\ge0$.
\end{enumerate}
\end{claim}

\begin{proof}
(a) Suppose that $y\in B$ in an iteration of the while loop during the processing of $y$; then $x=y$ by Line~\ref{line:partner}.
Since $y\in B=A\cup C$ and $y\notin A$ by the loop condition, we have $y\in C\subseteq B_0\subseteq H$.
Thus, $C-y\subseteq H-y$, and hence
\[
\lambda(y)=g(y\mid C-y)\ge g(y\mid H-y)=\gamma(y)\ge\Theta,
\]
where the first inequality follows from the submodularity of $g$ (Lemma~\ref{lem:g}), and the last inequality follows from Claim~\ref{clm:window}.
Therefore, $\hla(y)\ge\lambda(y)/(1+\eps)\ge\Theta/(1+\eps)>\Theta/(1+\eps)^2$, and hence the test in Line~\ref{line:test} succeeds for $x=y$, and $y$ is protected in Line~\ref{line:protect}.
Consequently, an exchange is performed only in an iteration with $y\notin B$; since $x$ is chosen from $C\setminus A\subseteq B$ in Line~\ref{line:partner}, we also have $x\ne y$.

(b) Let $y\in E$, let $x=x_y$, and let $B$ be the basis just before the exchange that inserts $y$.
Since the exchange is performed during the processing of $y$ at threshold $\Theta(y)$, Claim~\ref{clm:window} gives $\Theta(y)\le\gamma(y)=a_y$.
Since the exchange is performed only when the test in Line~\ref{line:test} fails, we have $\hla(x)\le\Theta(y)/(1+\eps)^2$, and hence $b_y=\lambda(x)\le(1+\eps)\hla(x)\le\Theta(y)/(1+\eps)\le a_y/(1+\eps)$.
Since $x\in B$, we have
\[
\Delta_y=g(B-x+y)-g(B)=g(y\mid B-x)-g(x\mid B-x).
\]
Since $B=A\cup C\subseteq H$ and $y\notin B$ by (a), we have $B-x\subseteq H-y$; moreover, $C-x\subseteq B-x$.
By the submodularity of $g$, we obtain $g(y\mid B-x)\ge g(y\mid H-y)=a_y$ and $g(x\mid B-x)\le g(x\mid C-x)=b_y$, and hence $\Delta_y\ge a_y-b_y$.
Finally, $a_y-b_y\ge a_y-a_y/(1+\eps)=\eps a_y/(1+\eps)$, which is positive since $a_y\ge\Theta(y)>0$.

(c) A protection (Line~\ref{line:protect}) adds an element of $C$ to $A$, and hence does not change $B=A\cup C$.
Thus, $B$ changes only through exchanges, and since each $y\in E$ is inserted by exactly one exchange, we have $\Delta=g(B_1)-g(B_0)=\sum_{y\in E}\Delta_y$.
Since $\Delta_y>0$ for every $y\in E$ by (b), we have $\Delta\ge0$.
\end{proof}

The next claim upper-bounds $D_1=z(H)-z(C)$ by tracking its change during the phase: each exchange changes it by at most $O(a_y)$, while increasing $g$ by at least $\Omega(\eps)\cdot a_y$ by \Cref{clm:exchange}(b).
The same argument also bounds $\sum_{y\in E}\Theta(y)$, since $\Theta(y)\le a_y$.
Recall that $\Theta(y)$ is the threshold at which $y$ is processed.

\begin{claim}\label{clm:charge}
At the end of the phase, both $D_1$ and $\sum_{y\in E}\Theta(y)$ are $O(\Delta/\eps)$.
\end{claim}

\begin{proof}
By Claim~\ref{clm:exchange}, $\Delta=\sum_{y\in E}\Delta_y\ge\frac{\eps}{1+\eps}\sum_{y\in E}a_y$, so $\sum_{y\in E}a_y=O(\Delta/\eps)$. Since $\Theta(y)\le a_y$ by Claim~\ref{clm:exchange}(b), this gives $\sum_{y\in E}\Theta(y)\le\sum_{y\in E}a_y=O(\Delta/\eps)$.

To show $D_1=O(\Delta/\eps)$, we track $z(H)-z(C)$ during the phase.
At the beginning of the phase, $z(H)-z(C)=0$ since $A=\emptyset$ and hence $H=C=B_0$.
The protections (Line~\ref{line:protect}) change neither $H$ nor $C$, since they only add an element of $C\subseteq B_0$ to $A$ and $H=B_0\cup A$.
Thus, $z(H)-z(C)$ changes only through exchanges.
Consider an exchange that inserts $y\in E$ and deletes $x$, and let $H$ and $C$ denote the sets just before it; the exchange replaces $H$ by $H+y$ and $C$ by $C-x$.
By~\eqref{eq:z-vs-g} and the submodularity of $g$, we have
\begin{align*}
z(H+y)-z(H)&\le g(y\mid H)\le g(y\mid H-y)=a_y,\\
z(C)-z(C-x)&\le g(x\mid C-x)=b_y\le a_y,
\end{align*}
where the last inequality follows from Claim~\ref{clm:exchange}(b).
Hence, each exchange increases $z(H)-z(C)$ by at most $2a_y$.
Since each $y\in E$ is inserted by exactly one exchange, we obtain $D_1=z(H)-z(C)\le 2\sum_{y\in E}a_y=O(\Delta/\eps)$ at the end of the phase.
\end{proof}

Using \Cref{obs:time}, we show that every protected element stays in $C$ and keeps a large loss.

\begin{claim}\label{clm:fixed}
Every protected element $x\in P$ belongs to the final $C$. Moreover, if $x$ was protected at threshold $\Theta$, then $\Theta\le(1+O(\eps))\cdot\lambda(x)$ for the final $C$.
\end{claim}

\begin{proof}
Let $x\in P$. When $x$ is protected (Line~\ref{line:protect}), $x$ belongs to $C$ and is added to $A$.
Since an element is deleted from $C$ only if it belongs to $C\setminus A$, and $x$ stays in $A$, the element $x$ is never deleted; hence $x$ belongs to the final $C$.

Next, suppose that $x$ was protected at threshold $\Theta$.
At that time, we have $\lambda(x)\ge\hla(x)>\Theta/(1+\eps)^2$.
Since $\lambda(x)$ does not decrease afterwards (Observation~\ref{obs:time}), we obtain $\Theta<(1+\eps)^2\cdot\lambda(x)\le(1+O(\eps))\cdot\lambda(x)$ for the final $C$.
\end{proof}

The next claim is the heart of the analysis. It plays the role of the approximate local optimality~\cite[Inequality~(3.4)]{buchbinder2025deterministic} in the analysis of Buchbinder and Feldman, and is the main tool for upper-bounding $D_2$.

\begin{claim}\label{clm:order}
Let $\theta_1\ge\theta_2\ge\dots\ge\theta_{|A|}$ be the thresholds at which the elements of the final $A$ entered $A$, listed in order of entry, and set $\theta_i=\Tmin$ for $|A|<i\le r$. Then, at the end of the phase, every $S\in\IL$ satisfies
\[
\sum_{y\in S}\gamma(y)\le(1+O(\eps))\sum_{i=1}^{r}\theta_i.
\]
\end{claim}

The proof of \Cref{clm:order} compares the order in which the elements of $S$ are processed with the order in which elements enter $A$.
Let $y_i$ be the $i$-th processed element of $S$, and consider the moment when its processing ends.
At this moment, each of $y_1,\dots,y_i$ is spanned by $A$, and hence $|A|\ge i$, i.e., the $i$-th element to enter $A$ did so no later than this moment.
Since thresholds decrease over time, this gives $\theta_i\ge\Theta(y_i)$.
On the other hand, \Cref{clm:window} gives $\gamma(y_i)\le(1+O(\eps))\cdot\Theta(y_i)$ at the end of the phase.
Combining these two inequalities and summing over $i$ yields the inequality of \Cref{clm:order}.

\begin{proof}[Proof of \Cref{clm:order}]
We say that $y$ is \emph{spanned} by $A$ if $y\in A$ or $A+y\notin\IL$.
The processing of an element $y$ ends only when the loop condition fails (i.e., $y$ is spanned by $A$) or after an exchange (Line~\ref{line:insert}) that inserts it into $A$, so every processed element is spanned by $A$ at the moment its processing ends.
It remains spanned afterwards, because $A$ only grows.

Let $y_1,\dots,y_k$ be the elements of $S$ that are processed during the phase (possibly $k<|S|$), in the order in which they are processed.
Fix $i\in[k]$.
When the processing of $y_i$ ends, the set $\{y_1,\dots,y_i\}\subseteq S$ is independent (as $S\in\IL$) and all of its elements are spanned by $A$.
Since $A\subseteq B$ is independent, if $|A|<i$ at that moment, then the augmentation property gives some $y_h\notin A$ ($h\le i$) with $A+y_h\in\IL$, which contradicts that $y_h$ is spanned by $A$.
Hence, $|A|\ge i$ when the processing of $y_i$ ends.
Since $A$ never shrinks, the $i$-th element to enter $A$ did so no later than the end of the processing of $y_i$.
Since thresholds never increase over time, $\theta_i\ge\Theta(y_i)$.
Therefore, at the end of the phase,
$
\gamma(y_i)\le(1+O(\eps))\cdot\Theta(y_i)\le(1+O(\eps))\cdot\theta_i,
$
where the first inequality holds since $\gamma(y_i)\le(1+O(\eps))\cdot\Theta(y_i)$ while $y_i$ is processed (Claim~\ref{clm:window}) and $\gamma(y_i)$ does not increase afterwards (Observation~\ref{obs:time}).

Let $y_{k+1},\dots,y_{|S|}$ be the remaining elements of $S$, which are unprocessed, in arbitrary order; note that $|S|\le r$ since $S\in\IL$.
For each $k<i\le|S|$, Claim~\ref{clm:window} gives $\gamma(y_i)\le(1+O(\eps))\cdot\Tmin\le(1+O(\eps))\cdot\theta_i$ at the end of the phase, where the last inequality holds since $\theta_i\ge\Tmin$ (every threshold is at least $\Tmin$).
Combining these bounds and using $\theta_i\ge0$ for every $i$, we obtain
$\ds
\sum_{y\in S}\gamma(y)=\sum_{i=1}^{|S|}\gamma(y_i)\le(1+O(\eps))\sum_{i=1}^{|S|}\theta_i\le(1+O(\eps))\sum_{i=1}^{r}\theta_i.
$
\end{proof}

Using \Cref{clm:order}, we obtain the following inequality, which is used to upper-bound $D_2$.

\begin{claim}\label{clm:opt}
At the end of the phase,
\[
\frac1\ell\sum_{j=1}^{\ell}\sum_{u\in\OPT}\gamma\bigl((u,j)\bigr)\le(1+O(\eps))\sum_{x\in C}\lambda(x)+O\Bigl(\sum_{y\in E}\Theta(y)+\fmax\Bigr).
\]
\end{claim}

\begin{proof}
For each $j\in[\ell]$, we have $\OPT\times\{j\}\in\IL$, and hence Claim~\ref{clm:order} with $S=\OPT\times\{j\}$ gives
\[
\sum_{u\in\OPT}\gamma\bigl((u,j)\bigr)\le(1+O(\eps))\sum_{i=1}^{r}\theta_i.
\]
Each element of $A$ entered $A$ either as a protected element $x\in P$, at a threshold of at most $(1+O(\eps))\cdot\lambda(x)$ by Claim~\ref{clm:fixed}, or as an inserted element $y\in E$, at threshold $\Theta(y)$.
The remaining $r-|A|$ values $\theta_i$ are equal to $\Tmin$, and they sum to at most $r\Tmin=\fmax$.
Hence,
\begin{align*}
\sum_{i=1}^{r}\theta_i
\le(1+O(\eps))\sum_{x\in P}\lambda(x)+\sum_{y\in E}\Theta(y)+\fmax
\le(1+O(\eps))\sum_{x\in C}\lambda(x)+\sum_{y\in E}\Theta(y)+\fmax,
\end{align*}
where the last inequality holds since $P\subseteq C$ by Claim~\ref{clm:fixed} and $\lambda(x)\ge0$ for every $x\in C\setminus P$ by the monotonicity of $g$.
Combining this with the first inequality, we have, for every $j\in[\ell]$,
\[
\sum_{u\in\OPT}\gamma\bigl((u,j)\bigr)
\le(1+O(\eps))\sum_{x\in C}\lambda(x)+O\Bigl(\sum_{y\in E}\Theta(y)+\fmax\Bigr).
\]
Since the right-hand side does not depend on $j$, summing this inequality over $j\in[\ell]$ and dividing by $\ell$ yields the claim.
\end{proof}

We now prove Lemma~\ref{lem:phase} using the claims established above.

\begin{proof}[Proof of Lemma~\ref{lem:phase}]
We have $g(B_1)-g(B_0)=\Delta\ge0$ by Claim~\ref{clm:exchange}(c). 
We apply Corollary~\ref{cor:compare} to the final sets $C\in\IL$ (recall that $C\subseteq B_0$) and $H$. In the notation of this section, $D_2=\frac1\ell\sum_{j}\sum_{u\in\OPT}\gamma((u,j))-\sum_{x\in C}\lambda(x)$. By Claims~\ref{clm:charge} and~\ref{clm:opt},
\[
D_1+D_2\le O(\eps)\sum_{x\in C}\lambda(x)+O\Bigl(\frac{\Delta}{\eps}\Bigr)+O(\fmax).
\]
By Lemma~\ref{lem:del-add}(a) and $z(C)\ge0$, we have $\sum_{x\in C}\lambda(x)\le\kappa\cdot f(\pi(C))$. 
Plugging these bounds into Corollary~\ref{cor:compare} and dividing by $\kappa$, we obtain
\[
\Bigl(1-\frac{\ell}{\kappa}\Bigr)\cdot f(\OPT)\le(1+O(\eps))\cdot f\bigl(\pi(C)\bigr)+O\Bigl(\frac{\Delta}{\eps\kappa}\Bigr)+O\Bigl(\frac{\fmax}{\kappa}\Bigr).
\]
By Fact~\ref{fact:coef}(d), we have $1-\ell/\kappa\ge1-1/e-\eps$ and $\kappa\ge\ell\ge1/\eps$. Together with $\fmax\le f(\OPT)$, this gives
\[
\Bigl(1-\frac1e-O(\eps)\Bigr)\cdot f(\OPT)\le(1+O(\eps))\cdot f\bigl(\pi(C)\bigr)+O(\Delta).
\]
Dividing by $1+O(\eps)$, we obtain $f(\pi(C))\ge(1-1/e-O(\eps))\cdot f(\OPT)-O(\Delta)$. Finally, $C\subseteq B_0$ and monotonicity of $f$ give $f(\pi(B_0))\ge f(\pi(C))$. 
Therefore, we obtain
\[
f\bigl(\pi(B_0)\bigr)\ge f\bigl(\pi(C)\bigr)\ge\Bigl(1-\frac1e-O(\eps)\Bigr)\cdot f(\OPT)-O(\Delta).
\]
\end{proof}

\section{Proof of the Main Theorem}\label{sec:main}

In this section, we present the approximation guarantee and the query complexity of Algorithm~\ref{alg:main} (Theorem~\ref{thm:core}), which implies Theorem~\ref{thm:main_r}.

\begin{theorem}\label{thm:core}
Algorithm~\ref{alg:main} returns a basis $S$ of $\M$ such that, w.h.p., $f(S)\ge(1-1/e-O(\eps))\cdot f(\OPT)$.
It uses $O(\eps^{-8}n\log^2(n/\eps))$ value oracle queries and $O(\eps^{-4}n+\eps^{-3}r\log r)$ independence oracle queries.
\end{theorem}

\begin{proof}
Since $\pi$ maps bases of $\ML$ to bases of $\M$, the output is a basis of $\M$.

\subparagraph*{Probability.}
By Lemma~\ref{lem:phase-queries}, regardless of the outcomes of the estimates, the algorithm calls $\Est$ $O(T\eps^{-2}n\log(n/\eps))=\Poly(n/\eps)$ times. By Lemma~\ref{lem:estimate}, each call is good w.h.p., even conditioned on the previous calls. Hence, by the union bound over the calls, w.h.p.\ all estimates are good. In the rest of the proof, we assume this.

\subparagraph*{Approximation Guarantee.}
Let $\Delta_t=g(B^{(t+1)})-g(B^{(t)})$, which is non-negative by Lemma~\ref{lem:phase}.
We have
\[
\sum_{t=0}^{T-1}\Delta_t=g(B^{(T)})-g(B^{(0)})\le g(B^{(T)})\le O(\ell^2)\cdot f(\OPT),
\]
where the last inequality follows from Observation~\ref{obs:range} and $\pi(B^{(T)})\in\I$.
Since $T=\ell^3$, some phase $t$ satisfies $\Delta_t\le O(\ell^2/T)\cdot f(\OPT)=O(\eps)\cdot f(\OPT)$. By Lemma~\ref{lem:phase} applied to this phase,
\[
f\bigl(\pi(B^{(t)})\bigr)\ge\Bigl(1-\frac1e-O(\eps)\Bigr)\cdot f(\OPT)-O(\Delta_t)=\Bigl(1-\frac1e-O(\eps)\Bigr)\cdot f(\OPT).
\]
Since the algorithm returns the best starting basis, the output also satisfies this bound.

\subparagraph*{Query Complexity.}
By Lemma~\ref{lem:phase-queries} and $T=O(\eps^{-3})$, the $T$ phases use $O(\eps^{-8}n\log^2(n/\eps))$ value oracle queries and $O(\eps^{-4}n+\eps^{-3}r\log r)$ independence oracle queries in total. The preprocessing uses $O(n)$ value and independence oracle queries, and choosing the best starting basis uses $T$ value oracle queries.
\end{proof}


\subsection{Deterministic Algorithm.} 
We obtain Theorem~\ref{thm:main_d} by replacing every call $\Est$ in Algorithm~\ref{alg:main} with the exact computation of $g(x\mid X-x)$ in \Cref{lem:exact}; all other steps are deterministic.
Since exact values are always good, the proof of Theorem~\ref{thm:core} applies without the probability argument, and the number of independence oracle queries does not change.
Each of the $O(T\eps^{-2}n\log(n/\eps))$ calls to $\Est$ now uses $2^{\ell}=2^{O(1/\eps)}$ value oracle queries, and hence the total number of value oracle queries is $O\bigl(T\eps^{-2}n\log(n/\eps)\bigr)\cdot2^{O(1/\eps)}=O\bigl(n\log n\cdot2^{O(1/\eps)}\bigr)$.

\paragraph*{Declaration of Generative AI Use.}
The main ideas behind our protected local search algorithm and its analysis were developed largely with the assistance of GPT-6 Astra (OpenAI).
We also used GPT-6 Astra (OpenAI), Claude Opus 5.5, and Claude Fable 5.1 (Anthropic) to assist in drafting the manuscript and improving its language and presentation.
We carefully verified and revised the proofs and take full responsibility for the paper.

\bibliography{biblio}

\newpage

\appendix

\section{Omitted Proofs}\label{sec:omitted}

In this appendix, we restate the lemmas whose proofs were omitted from the main body, and provide their proofs.

\subsection{Omitted Proofs in \Cref{sec:potential}} \label{appendix:potential}

\coef*

\begin{proof}[Proof of Fact~\ref{fact:coef}]
(a) $\alpha_{k+1}/\alpha_k=(1+1/\ell)\binom{\ell-1}{k-1}\big/\binom{\ell-1}{k}=(1+1/\ell)\,k/(\ell-k)$.

(b) There are $\binom{\ell-1}{k-1}$ sets $J\ni i$ of size $k$ in the sum, and $\binom{\ell-1}{k-1}\alpha_k=(1+1/\ell)^{k-1}$. Hence, the sum is $\sum_{k=1}^{\ell}(1+1/\ell)^{k-1}=\ell\bigl((1+1/\ell)^{\ell}-1\bigr)=\kappa-\ell$. For the second identity, there are $\binom{\ell}{k}$ sets $J$ of size $k$ in the sum, and $\binom{\ell}{k}\frac{k}{\ell} = \binom{\ell-1}{k-1}$.
Hence, the sum equals $\sum_{k=1}^{\ell}\binom{\ell-1}{k-1}\alpha_k=\sum_{k=1}^{\ell}(1+1/\ell)^{k-1}$, which is $\kappa-\ell$ as shown above.

(c) There are $\binom{\ell}{k}$ sets of size $k$ in the sum. Since $\binom{\ell}{k} = \frac{\ell}{k}\binom{\ell-1}{k-1}$, we have $\binom{\ell}{k}\alpha_k=\frac{\ell}{k}(1+1/\ell)^{k-1} \le \ell (1+1/\ell)^{k-1} \le e\ell$, where the last inequality uses $(1+1/\ell)^{k-1} \le (1+1/\ell)^{\ell} \le e$.
Summing over $k\in[\ell]$ gives the claim.

(d) For every $\ell > 0$, we have $(1+1/\ell)^{\ell}\le e\le(1+1/\ell)^{\ell+1}$. The first inequality gives $\kappa\le e\ell$, and the second gives $\ell/\kappa=(1+1/\ell)^{-\ell}\le(1+1/\ell)/e\le1/e+1/\ell$.
\end{proof}

\marginal*

\begin{proof}[Proof of Lemma~\ref{obs:marginal}]
By the definition of $g$,
\[
g(x\mid X-x) = g(X+x)-g(X-x) = \sum_{J\subseteq[\ell]}\alpha_{|J|}\Bigl(f\bigl(\pi_J(X+x)\bigr)-f\bigl(\pi_J(X-x)\bigr)\Bigr).
\]
We consider the value $f\bigl(\pi_J(X+x)\bigr)-f\bigl(\pi_J(X-x)\bigr)$.
If $i\notin J$, then $\pi_J(X+x)=\pi_J(X-x)$, and hence the term is zero.
If $i\in J$, then $\pi_J(X+x)=\pi_J(X-x)+u$, and hence
$
f\bigl(\pi_J(X+x)\bigr)-f\bigl(\pi_J(X-x)\bigr)=f\bigl(u\mid\pi_J(X-x)\bigr).
$
Combining these yields the claim.
\end{proof}

\range*

\begin{proof}[Proof of Observation~\ref{obs:range}]
The first inequality holds since
\[
g(X)=\sum_{J\subseteq[\ell]}\alpha_{|J|}\cdot f\bigl(\pi_J(X)\bigr)
\le \Bigl(\sum_{J\subseteq[\ell]}\alpha_{|J|}\Bigr)\cdot f\bigl(\pi(X)\bigr)
= O(\ell^2)\cdot f\bigl(\pi(X)\bigr),
\]
where the inequality follows from $\alpha_{|J|}\ge 0$ and $f(\pi_J(X))\le f(\pi(X))$ (by $\pi_J(X)\subseteq\pi(X)$ and the monotonicity of $f$), and the equality follows from Fact~\ref{fact:coef}(c).
For the second inequality, let $y=(u,i)$. Then
\[
g(y\mid X)\le g(y\mid\emptyset)
=\sum_{J\ni i}\alpha_{|J|}\cdot f(u\mid\emptyset)
=(\kappa-\ell)\cdot f(u\mid\emptyset)
\le(\kappa-\ell)\cdot f(\{u\}),
\]
where the first inequality follows from the submodularity of $g$ (Lemma~\ref{lem:g}), the first equality follows from Observation~\ref{obs:marginal} and $\pi_J(\emptyset)=\emptyset$, the second equality follows from Fact~\ref{fact:coef}(b), and the last inequality follows from $f(\emptyset)\ge 0$.
\end{proof}

\subsection{Omitted Proofs in \Cref{subsec:comp}: Comparison Inequality} \label{appendix:deladd}

\deladd*

\begin{proof}[Proof of \Cref{lem:del-add}]
(a) We have
\begin{align}
\sum_{x\in C}g(x\mid C-x)
&=\sum_{(u,i)\in C}\ \sum_{J\ni i}\alpha_{|J|}\cdot f\bigl(u\mid\pi_J(C)-u\bigr) \notag \\
&=\sum_{J\subseteq[\ell]}\alpha_{|J|}\sum_{i\in J}\ \sum_{u\in\pi_{\{i\}}(C)}f\bigl(u\mid\pi_J(C)-u\bigr) \notag\\
&\le\sum_{J\subseteq[\ell]}\alpha_{|J|}\sum_{i\in J}\Bigl(f\bigl(\pi_J(C)\bigr)-f\bigl(\pi_{J-i}(C)\bigr)\Bigr) \label{eq:del-bound},
\end{align}
where the first equality follows from Observation~\ref{obs:marginal} and $\pi_J(C-x)=\pi_J(C)-u$ for $x=(u,i)\in C$ and $J\ni i$, the second equality follows by exchanging the order of summation, and the inequality follows from Fact~\ref{fact:submod}(a) with $T=\pi_J(C)$ and $S=\pi_{J-i}(C)$, since $\pi_{\{i\}}(C)=\pi_J(C)\setminus\pi_{J-i}(C)$.
Here, both $\pi_J(C-x)=\pi_J(C)-u$ and $\pi_{\{i\}}(C)=\pi_J(C)\setminus\pi_{J-i}(C)$ hold because $C$ contains at most one copy of each element.

The right-hand side of the inequality (\ref{eq:del-bound}) plus $z(C) := \sum_{J\subseteq[\ell]}\frac{|J|}{\ell}\,\alpha_{|J|} \cdot f\bigl(\pi_J(C)\bigr)$ equals
\[
\sum_{J \subseteq [\ell]}|J|\,\alpha_{|J|}\,f\bigl(\pi_J(C)\bigr)
-\sum_{J \subseteq [\ell]}\alpha_{|J|}\sum_{i\in J}f\bigl(\pi_{J-i}(C)\bigr)
+\sum_{J \subseteq [\ell]}\frac{|J|}{\ell}\,\alpha_{|J|}\,f\bigl(\pi_J(C)\bigr).
\]
The first and third sums together equal $\sum_{K}|K|(1+1/\ell)\,\alpha_{|K|}\,f(\pi_K(C))$.
For the second sum, the pairs $(J,i)$ with $i\in J$ correspond one-to-one to the pairs $(K,i)$ with $i\notin K$ via $K=J-i$, and hence
\[
\sum_{J}\alpha_{|J|}\sum_{i\in J}f\bigl(\pi_{J-i}(C)\bigr)
=\sum_{K}\sum_{i\notin K}\alpha_{|K|+1}\,f\bigl(\pi_K(C)\bigr)
=\sum_{K}(\ell-|K|)\,\alpha_{|K|+1}\,f\bigl(\pi_K(C)\bigr).
\]
Therefore, the right-hand side of the inequality (\ref{eq:del-bound}) plus $z(C)$ equals $\sum_{K}c_K\cdot f(\pi_K(C))$ with
\[
c_K=|K|\Bigl(1+\frac{1}{\ell}\Bigr)\alpha_{|K|}-(\ell-|K|)\,\alpha_{|K|+1}.
\]
Setting $\alpha_{\ell+1}=0$, we obtain
\[
\sum_{x\in C}g(x\mid C-x)+z(C)\le\sum_{K\subseteq[\ell]}\Bigl(|K|(1+1/\ell)\alpha_{|K|}-(\ell-|K|)\alpha_{|K|+1}\Bigr)\cdot f\bigl(\pi_K(C)\bigr).
\]
By Fact~\ref{fact:coef}(a), the coefficient vanishes for $1\le|K|<\ell$. For $K=\emptyset$, the coefficient is negative. For $K=[\ell]$, the coefficient is $(\ell+1)\alpha_\ell=\kappa$, which completes the proof of (a).

(b) Fix $j\in[\ell]$ and $u\in\OPT$. 
We have
\[
g\bigl((u,j)\bigm| H-(u,j)\bigr)
=\sum_{J\ni j}\alpha_{|J|}\cdot f\bigl(u\bigm|\pi_J(H-(u,j))\bigr)
\ge\sum_{J\ni j}\alpha_{|J|}\cdot f\bigl(u\bigm|\pi_J(H)\bigr),
\]
where the equality follows from Observation~\ref{obs:marginal}.
For the inequality, if $(u,j)\notin H$, then $H-(u,j)=H$, and hence it holds with equality.
If $(u,j)\in H$, then $u\in\pi_J(H)$ for every $J\ni j$, and hence the right-hand side is $0$, while the left-hand side is non-negative by the monotonicity of $g$ (Lemma~\ref{lem:g}).
We also have 
\[
\sum_{u\in\OPT}f\bigl(u\bigm|\pi_J(H)\bigr)
\ge f\bigl(\OPT\cup\pi_J(H)\bigr)-f\bigl(\pi_J(H)\bigr)
\ge f(\OPT)-f\bigl(\pi_J(H)\bigr),
\]
where the first inequality follows from Fact~\ref{fact:submod}(b), and the second follows from the monotonicity of $f$.
Combining the two inequalities above, for each $j\in[\ell]$,
\[
\sum_{u\in\OPT}g\bigl((u,j)\bigm| H-(u,j)\bigr)
\ge\sum_{J\ni j}\alpha_{|J|}\Bigl(f(\OPT)-f\bigl(\pi_J(H)\bigr)\Bigr).
\]
Summing this inequality over $j\in[\ell]$ and dividing by $\ell$, we obtain
\begin{align*}
\frac1\ell\sum_{j=1}^{\ell}\sum_{u\in\OPT}g\bigl((u,j)\bigm| H-(u,j)\bigr)
&\ge\frac1\ell\sum_{j=1}^{\ell}\sum_{J\ni j}\alpha_{|J|}\Bigl(f(\OPT)-f\bigl(\pi_J(H)\bigr)\Bigr)\\
&=\sum_{J\subseteq[\ell]}\frac{|J|}{\ell}\,\alpha_{|J|}\Bigl(f(\OPT)-f\bigl(\pi_J(H)\bigr)\Bigr)\\
&=(\kappa-\ell)\cdot f(\OPT)-z(H),
\end{align*}
where the first equality holds since each $J\subseteq[\ell]$ appears exactly $|J|$ times in the double sum, once for each $j\in J$, and the second equality follows from Fact~\ref{fact:coef}(b) and the definition of $z$.\qedhere
\end{proof}

\subsection{Omitted Proofs in \Cref{sec:estimation}: Estimating {$g(x\mid X-x)$}} \label{appendix:estimate}

\est*

To prove Lemma~\ref{lem:estimate}, we first show the following lemma.

\begin{lemma}[Sampler]\label{lem:sampler}
Let $x=(u,i)\in\NL$ and $X\subseteq\NL$. Choose $k\in[\ell]$ uniformly at random, then choose $J$ uniformly at random among the subsets of $[\ell]$ of size $k$ that contain $i$, and let
\[
Y=\ell(1+1/\ell)^{k-1}\cdot f\bigl(u\mid\pi_J(X-x)\bigr).
\]
Then, $\E[Y]=g(x\mid X-x)$ and $0\le Y\le e\ell\cdot\E[Y]$.
\end{lemma}

\begin{proof}
Given $|J|=k$, the set $J$ is uniformly distributed over the $\binom{\ell-1}{k-1}$ sets of size $k$ containing $i$, and $\Pr[|J|=k]=1/\ell$.
Hence, each set $J\ni i$ is chosen with probability $1/(\ell\binom{\ell-1}{|J|-1})$.
Since $\binom{\ell-1}{k-1}\alpha_k=(1+1/\ell)^{k-1}$ for every $k\in[\ell]$, we obtain
\[
\E[Y]
=\sum_{J\ni i}\frac{\ell\,(1+1/\ell)^{|J|-1}}{\ell\binom{\ell-1}{|J|-1}}\cdot f\bigl(u\mid\pi_J(X-x)\bigr)
=\sum_{J\ni i}\alpha_{|J|}\cdot f\bigl(u\mid\pi_J(X-x)\bigr)
=g(x\mid X-x),
\]
where the last equality follows from Observation~\ref{obs:marginal}.

Let $h_0=f(u\mid\pi_{\{i\}}(X-x))$. Every sampled $J$ contains $i$, so $\pi_J(X-x)\supseteq\pi_{\{i\}}(X-x)$, and submodularity gives $Y\le\ell(1+1/\ell)^{\ell-1}h_0\le e\ell h_0$.
On the other hand, $J=\{i\}$ with probability $1/\ell$, and then $Y=\ell h_0$. Since $Y\ge0$, this gives $\E[Y]\ge \frac{1}{\ell} \cdot \ell h_0 = h_0$, and hence $Y\le e\ell\cdot\E[Y]$.
\end{proof}

We use the following standard Chernoff bound.

\begin{fact}[Chernoff bound; see, e.g., {\cite[Theorem~1.1]{dubhashi2009concentration}}]\label{fact:chernoff}
Let $Y_1,\dots,Y_m$ be independent random variables taking values in $[0,R]$, and let $\mu=\E[\sum_{s=1}^mY_s]$. Then, for $0<\eta<1$,
\[
\Pr\Bigl[\Bigl|\sum_{s=1}^mY_s-\mu\Bigr|\ge\eta\mu\Bigr]\le2\exp\Bigl(-\frac{\eta^2\mu}{3R}\Bigr).
\]
\end{fact}

\begin{algorithm}[t]
\caption{$\Est(x,X)$}\label{alg:estimate}
\begin{algorithmic}[1]
\Require $x\in\NL$ and $X\subseteq\NL$
\State draw $m=\Theta(\ell\eps^{-2}\log(n/\eps))$ independent samples $Y_1,\dots,Y_m$ of $Y$ in Lemma~\ref{lem:sampler}
\State \Return $\frac{1}{1+\eps/3}\cdot\frac1m\sum_{s=1}^{m}Y_s$
\end{algorithmic}
\end{algorithm}

\begin{proof}[Proof of Lemma~\ref{lem:estimate}]
We show that Algorithm~\ref{alg:estimate} satisfies the conditions of Lemma~\ref{lem:estimate}. 
Let $a=g(x\mid X-x)$. If $a=0$, then $Y=0$ with probability $1$ and the output is $0$. 
Otherwise, by Lemma~\ref{lem:sampler}, each sample $Y_s$ satisfies $0\le Y_s\le e\ell a$ and $\E[Y_s]=a$.
Hence, Fact~\ref{fact:chernoff} with $R=e\ell a$, $\mu=ma$, and $\eta=\eps/3$ shows that the average $\overline{Y}$ of the $m$ samples satisfies $|\overline{Y}-a|<\eps a/3$ with probability at least $1-2\exp(-\Omega(\eps^2m/\ell))$, which is w.h.p.\ by the choice of $m$. 
In this case, the output $\widehat g=\overline{Y}/(1+\eps/3)$ satisfies $\widehat g\le a$ and $\widehat g\ge\frac{1-\eps/3}{1+\eps/3}a\ge\frac{a}{1+\eps}$.
Each sample uses two value oracle queries, $f(\pi_J(X-x)+u)$ and $f(\pi_J(X-x))$. Since $\ell=O(1/\eps)$, the number of value oracle queries is $2m=O(\eps^{-3}\log(n/\eps))$.
\end{proof}

\end{document}